\documentstyle[12pt]{article}
\def\a{\alpha}      
\def\d{\delta}

 \def\om{\omega}  \def\Om{\Omega}

\def\kp{\xi^{(++)}}  \def\km{\xi^{(--)}}  \def\kpm{\xi^{(\pm\pm)}}
\def\M{\underline m}   \def\N{\underline n}
\def\A{\underline a}   \def\B{\underline b}   \def\C{\underline c}
\def\GL{G_{\cal L}}   \def\GR{G_{\cal R}}  \def\GLR{G_{\cal L,\cal R}}
\def\pap{\partial_{(++)}}   \def\pam{\partial_{(--)}}
\def\Mpp{M^{(++)}_{(++)}}   \def\Mmm{M^{(--)}_{(--)}}
\def\Mpm #1{M^{++ #1}_{(--)}}   \def\Mmp #1{M^{-- #1}_{(++)}}
\def\Mpmmp #1{M^{\pm\pm #1}_{(\mp\mp)}}

\def\semiprod{\subset\!\!\!\!\!\!\times}
\def\PRL #1 #2 #3{{\sl Phys. Rev. Lett.} {\bf#1} (#2) #3}
\def\NPB #1 #2 #3{{\sl Nucl. Phys.} {\bf B #1} (#2) #3}
\def\NPBFS #1 #2 #3 #4{{\sl Nucl. Phys.} {\bf B #2} [FS#1] (#3) #4}
\def\CMP #1 #2 #3{{\sl Commun. Math. Phys.} {\bf #1} (#2) #3}
\def\PRD #1 #2 #3{{\sl Phys. Rev.} {\bf D #1} (#2) #3}
\def\PLA #1 #2 #3{{\sl Phys. Lett.} {\bf #1A} (#2) #3}
\def\PLB #1 #2 #3{{\sl Phys. Lett.} {\bf B #1} (#2) #3}
\def\JMP #1 #2 #3{{\sl J. Math. Phys.} {\bf #1} (#2) #3}
\def\PTP #1 #2 #3{{\sl Prog. Theor. Phys.} {\bf #1} (#2) #3}
\def\SPTP #1 #2 #3{{\sl Suppl. Prog. Theor. Phys.} {\bf #1} (#2) #3}
\def\AoP #1 #2 #3{{\sl Ann. of Phys.} {\bf #1} (#2) #3}
\def\PNAS #1 #2 #3{{\sl Proc. Natl. Acad. Sci. USA} {\bf #1} (#2) #3}
\def\RMP #1 #2 #3{{\sl Rev. Mod. Phys.} {\bf #1} (#2) #3}
\def\PR #1 #2 #3{{\sl Phys. Reports} {\bf #1} (#2) #3}
\def\AoM #1 #2 #3{{\sl Ann. of Math.} {\bf #1} (#2) #3}
\def\UMN #1 #2 #3{{\sl Usp. Mat. Nauk} {\bf #1} (#2) #3}
\def\FAP #1 #2 #3{{\sl Funkt. Anal. Prilozheniya} {\bf #1} (#2) #3}
\def\FAaIA #1 #2 #3{{\sl Functional Analysis and Its Application} {\bf
#1} (#2) #3}
\def\BAMS #1 #2 #3{{\sl Bull. Am. Math. Soc.} {\bf #1} (#2)
#3} \def\TAMS #1 #2 #3{{\sl Trans. Am. Math. Soc.} {\bf #1} (#2) #3}
\def\InvM #1 #2 #3{{\sl Invent. Math.} {\bf #1} (#2) #3}
\def\LMP #1 #2 #3{{\sl Letters in Math. Phys.} {\bf #1} (#2) #3}
\def\IJMPA #1 #2 #3{{\sl Int. J. Mod. Phys.} {\bf A #1} (#2) #3}
\def\AdM #1 #2 #3{{\sl Advances in Math.} {\bf #1} (#2) #3}
\def\RMaP #1 #2 #3{{\sl Reports on Math. Phys.} {\bf #1} (#2) #3}
\def\IJM #1 #2 #3{{\sl Ill. J. Math.} {\bf #1} (#2) #3}
\def\APP #1 #2 #3{{\sl Acta Phys. Polon.} {\bf #1} (#2) #3}
\def\TMP #1 #2 #3{{\sl Theor. Mat. Phys.} {\bf #1} (#2) #3}
\def\JPA #1 #2 #3{{\sl J. Physics} {\bf A#1} (#2) #3}
\def\JSM #1 #2 #3{{\sl J. Soviet Math.} {\bf #1} (#2) #3}
\def\MPLA #1 #2 #3{{\sl Mod. Phys. Lett.} {\bf A #1} (#2) #3}
\def\JETP #1 #2 #3{{\sl Sov. Phys. JETP} {\bf #1} (#2) #3}
\def\JETPL #1 #2 #3{{\sl Sov. Phys. JETP Lett.} {\bf #1} (#2) #3}
\def\PHSA #1 #2 #3{{\sl Physica} {\bf A #1} (#2) #3}
\def\CQG #1 #2 #3{{\sl Class. Quantum Grav.} {\bf #1} (#2) #3}
\def\SJNP #1 #2 #3{{\sl Sov. J. Nucl. Phys. (Yadern.Fiz.)} {\bf #1} (#2) #3}

\newcommand{\p}[1]{(\ref{#1})}

\begin{document}
\renewcommand{\thefootnote}{\fnsymbol{footnote}}
\thispagestyle{empty}
\begin{flushright}
{\bf  
IC/98/94 \\
TUW--98--21
\\ hep-th/9810038} \\
\end{flushright}

\begin{center}
{\Large \bf
General Solution of   \\
String Inspired Nonlinear Equations}

\vskip 0.5cm

{\bf I. Bandos \footnote{
Lise Meitner Fellow of the 
''Fonds zur F\"{o}rderung der 
wissenschaftlichen Forschung''  
at the Institut f\"{u}r Theoretische Physik, 
Technische Universit\"{a}t Wien, A-1040 Vienna Austria, 
\\ 
E-mail: bandos@tph32.tuwien.ac.at}}

{\it The Abdus Salam ICTP, \\
P.O. Box 586, \\
34100, Trieste, Italy \\
and
\\
 Institute for Theoretical Physics, \\
     NSC Kharkov Institute of Physics and Technology, \\
     310108, Kharkov, Ukraine}

\vskip 0.5cm

{\bf E. Ivanov}\footnote{E-mail: eivanov@thsun1.jinr.ru}

{\it Bogoliubov Laboratory of Theoretical Physics,
JINR,\\
141 980 Dubna, Moscow Region, Russia} \\

\vskip 0.5cm

{\bf A. A. Kapustnikov \footnote{E-mail: alexandr@ff.dsu.dp.ua}}

{\it Department of Physics, Dnepropetrovsk University, \\
     320625, Dnepropetrovsk, Ukraine}

\vskip 0.5cm

{\bf S. A. Ulanov \footnote{E-mail: theorph@ff.dsu.dp.ua}}

{\it Department of Physics, Dnepropetrovsk University, \\
     320625, Dnepropetrovsk, Ukraine}
\end{center}

\setcounter{page}1
\renewcommand{\thefootnote}{\arabic{footnote}}
\setcounter{footnote}0
\begin{abstract}
We present the general solution of the system of
coupled nonlinear equations describing
dynamics of  $D$--dimensional bosonic string in
the geometric (or embedding) approach.
The solution is parametrized in terms of two sets of the left- and
right-moving Lorentz harmonic variables providing
a special coset space realization of the product of two
$(D-2)$ dimensional spheres
  $$
  S^{D-2}={SO(1,D-1) \over SO(1,1) \times SO(D-2) \semiprod
  K_{D-2}}~.
  $$
\end{abstract}

\section*{Introduction}

The bosonic string (and $p$-brane) theory allows a geometric
description in terms of extrinsic geometry of the worldsheet treated as
a surface embedded into a flat $D$-dimensional Minkowski space
\cite{geom,BNbook,bpstv,Ei}
\footnote
{See \cite{bpstv} for a supersymmetric generalization of the
classical surface theory and its application to superstrings and
$N=1$ superbranes.}.

In this approach the dynamics of a free relativistic
string is described by the Maurer-Cartan equation supplemented with
the additional conditions which insure the string
worldsheet to be a minimal surface embedded into the flat
target space-time.
All these additional constraints can be solved
algebraically, after which one is left with some $so(1,D-1)$ valued connection form
whose curvature vanishes due to the Maurer-Cartan equations \cite{zero}.
Thus, though the string in the geometric approach is described
by nonlinear equations \cite{geom,BNbook,zero}, the latter
are finally reduced to the zero curvature conditions for a $so(1,D-1)$ valued
connection form properly specified in terms of the independent field
variables \cite{zero}.

The system of independent equations describing a free string theory in
the Minkowski space of arbitrary dimension $D$ was
derived, for the first time, by Zheltukhin \cite{geom}.
In the form when all the gauge
symmetries inherent in the string theory are retained, this system
is formed by the WZNW sigma-model-type equation
  \begin{equation}\label{WZNW1}
  \pam \left((\pap G) G^T\right)^{ij} =
  e^{2W} G^{[i| k} M^{(++)k}_{(--)} M^{(--)|j]}_{(++)}
  \end{equation}
for the $SO(1,D-1)$ valued matrix field $G^{ij}$
  $$
  G G^T = I~,
  $$
and the Liouville-type equation
  \begin{equation}\label{Liouville0}
  \pap \pam W = {1 \over 4} M^{(--)i}_{(++)} G^{ij} M^{(++)j}_{(--)} e^{2W}
  \end{equation}
for a ``scalar density'' $e^W$ \cite{zero}. Thus the considered
system involves these two independent fields, as well as two
chiral $SO(D-2)$ vector fields $M^{(--)i}_{(++)}$,
$M^{(++)i}_{(++)}$~,
  \begin{equation}\label{chirality}
  \pam M_{(++)}^{(--)j} = 0, \qquad \pap M_{(--)}^{(++)j} = 0,
  \end{equation}
which appear in the right hand side.

Already at the early stage of developing the geometric approach
it was observed that at least for the low dimensions $D$ this system
is exactly solvable. Indeed, it is reduced to the Liouville equation
for $D = 3$ and a complex Liouville equation for $D=4$.
The general solution for $D = 5$ was found
in \cite{BNbook}.  However, for the generic case the
general solution of the equations \p{WZNW1}, \p{Liouville0}, \p{chirality}
was unknown so far.

In this paper we present the general solution of these equations for
any value of $D$.

\bigskip

First of all, our result gives a new example of non-trivial system of
exactly solvable nonlinear equations.

\bigskip

Secondly, this opens a possibility to study the classical and
quantum string theory in terms of new left- and right-moving
variables which parametrize the general solution.

\bigskip

The meaning and origin of our results require some comments.

\bigskip

The standard equations of motion for the string theory
become linear in the conformal gauge.
The general solution of these
linear equations is given by the sum of chiral functions (subjected to
the Virasoro constraints). The
string-inspired nonlinear equations (i.e. the equations describing
bosonic string theory in the geometric approach) encode
the information about just the same dynamical system. Thus it is natural to
expect (and, as we demonstrate here, this is indeed the case) that
these equations are exactly solvable
and that their general solution should be expressible in terms of
chiral data.

Thus, the first step was to seek for an adequate set of chiral variables
appropriate for constructing the general solution. It turned out that 
the necessary variables  were provided by the harmonic approach \cite{gikos} 
adapted to the case of Lorentz groups in refs. \cite{sok} - \cite{bzstr}.  

Namely,  in ref. \cite{Kapustn} it was found that
the ``constrained chiral twistor-like
variables'' (which can be identified with the spinor $SO(1,D-1)$ Lorentz
harmonics \cite{BandH,gds,bzstr}) can be used to obtain a
covariant solution of the chiral Virasoro constraints
  $$
  \pap X_L^{\underline{m}} \pap X_L^{\underline{m}} = 0,
\qquad \pam X_R^{\underline{m}} \pam X_R^{\underline{m}} = 0,
  $$ $$
  \pam X_L^{\underline{m}}=0, \qquad \pap X_R^{\underline{m}} = 0, \quad
(\underline{m} = 0,1 \dots D-1 )
  $$
which are to be imposed on the solutions $X_L^{\underline{m}}$,
$X_R^{\underline{m}}$ of the string equations of motion
in the conformal gauge. The $D=3,4,6$ strings were treated
in this way.
It was analysed how the fields composed from such chiral twistors are
related to the corresponding string-inspired nonlinear equations.

\bigskip

Here, instead of solving the chiral Virasoro constraints
of $D$-dimensional string theory, we
{\sl construct two chiral moving frames} given by the
two sets of chiral $SO(1,D-1)$ Lorentz harmonic variables \cite{sok},
and identify the left- and right-moving vectors
$\pap X_L^{\underline{m}}$ and
$\pam X_R^{\underline{m}}$ with the light-like components of the
relevant harmonic matrix.
Each set of the moving frame variables parametrizes a special coset space
of the group $SO(1,D-1)$. Despite the fact that the group
$SO(1,D-1)$ on its own is non-compact, this coset space is compact and
is isomorphic to the $D$-dimensional sphere \cite{gds}
  $$
  S^{D-2}= SO(1,D-1)/[SO(1,1) \times SO(D-2) \semiprod K_{D-2}]~.
  $$
It turns out to provide us with the appropriate chiral data for constructing
the general solution of the string-inspired nonlinear equations
\p{WZNW1} -- \p{chirality}. We construct such a solution explicitly
and argue that this is the {\sl general} one.

\subsection*{ Basic notation}

Our conventions basically coincide with those of Refs. \cite{zero}.
The indices
$m,n,... = 0,1 = (++), (--)$ ($v^{m} \equiv v^{\pm\pm}= (v^0 \pm v^1)$)
label the worldsheet vectors, while $ {\M}, {\N},... = 0,1,...,(D-1)$ are
the flat target space vector indices.

For the tangent space indices we use the notations
$a,b,... = 0,1 $ and
$\A, \B,... = 0,1,...,(D-1)$.
Thus, in all cases, underlined indices correspond to the $D$-dimensional
target space and the non-underlined ones refer to the two-dimensional ($d=2$)
worldsheet.

The indices $+, -$ denote the weights of tangent space vectors
and spinors with respect to both the worldsheet Lorentz group $SO(1,1)$
and the $SO(1,1)$ subgroup of the target space Lorentz group
$SO(1,D-1)$. These two subgroups are identified with each
other in the version of the
``geometric approach'' to string theory which forms the basis of the 
present consideration (see \cite{bpstv,zero} and refs.
therein). E.g., we write
$V^a = (V^0, V^1) = ( {1 \over 2} (V^{++} + V^{--}),
{1 \over 2} (V^{++} - V^{--}))$ for the $d=2$ tangent space vectors
(and reserve the notations $\psi^\a= (\psi^+,\psi^- ) $
for the $d=2$ spinors which appear in supersymmetric generalizations
of our approach).
The indices $+, -$ within the parentheses
denote the weights with respect to
the $d=2$ conformal symmetry, e.g.
$d\xi^m = (d\xi^0, d\xi^1) = ( {1 \over 2}
(d\xi^{(++)} + d\xi^{(--)}), {1 \over 2} (d\xi^{(++)} - d\xi^{(--)}))$.
The weights with respect to the chiral affine $SO(1,1)_L$ and
$SO(1,1)_R$ transformations are indicated in the same way.

We use the subscripts $L$ and $R$ to denote the chiral functions
of the string worldsheet coordinates $\xi^m = (\kp, \km)$
  $$
  f_L = f_L (\kp), \qquad f_R = f_R (\km),
  $$ $$
  \pam f_L = 0, \qquad \pap f_R = 0~.
  $$
They should not be confused with the calligraphic subscripts
${\cal L}$ and ${\cal R}$ carried by some fields.
The chiral  Lorentz harmonics (chiral moving frame variables)
are denoted by the letters $l$ and $r$, while the generic (non-chiral)
Lorentz harmonics are denoted by $u$.

\section{Geometric approach to string dynamics
and \break string--inspired nonlinear equations}

\subsection{Lorentz harmonics}

We begin with the definition of the moving frame variables
(Lorentz harmonics \cite{gikos,sok}) which are the basic ingredients
of the geometric approach to $D$-dimensional bosonic string theory
\cite{bzstr,bpstv,zero}
\begin{equation}\label{harm}
u^{~\A }_{\M}(\xi)\equiv
\left({1\over 2} (u^{++}_{\underline{m}}+ u^{--}_{\underline{m}}),
\; u^{i}_{\underline{m}},\; {1\over 2} (u^{++}_{\underline{m}}-
u^{--}_{\underline{m}}) \right)
  \end{equation}
$$
 \M =0,1,...,D-1,  \qquad \A=0,1,...,D-1~.
$$
These objects are subject to the following orthonormality
conditions
  \begin{equation}\label{ortho}
  u^{\M}_{~\A} u_{\M\B} = \eta_{\A\B} \equiv \mbox{diag} (1,-1,...,-1),
  \end{equation}
$$
~~~~~~~~~~~~~~~~~~~~~~~~\Leftrightarrow \cases{
u_{\M}^{++}  u^{\M ++} = 0, ~~~  u_{\M}^{--}  u^{\M --} = 0, ~~~\cr
u_{\M}^{++}  u^{\M --} = 2, ~~~ \cr
u_{\M}^{++}  u^{\M i} = 0, ~~~~ u_{\M}^{--}  u^{\M i} = 0, ~~~\cr
u_{\M}^{i}  u^{\M j} = - \d^{ij}. ~~~\cr
}
$$
They  imply that the $D\times D$ matrix
$  u^{~\A}_{\M}$ \p{harm} belongs to the group $SO(1,D-1)$
  \begin{equation}\label{harm1}
u^{~\A }_{\M}(\xi)
\qquad \in \qquad SO(1,D-1)~.
  \end{equation}
The completeness condition
  \begin{equation}\label{compl}
  u^{~\A}_{\M} u_{\N\A} \equiv
{1 \over 2}  u^{++}_{\M} u^{--}_{\N}  + {1 \over 2}  u^{--}_{\M} u^{++}_{\N}
-  u^{i}_{\M} u^{i}_{\N} =
\eta_{\M\N} \equiv \mbox{diag} (1,-1,...,-1)
  \end{equation}
follows from Eq. \p{ortho}.
In Eqs. \p{harm}, \p{compl} the light-like notation  is used
  \begin{eqnarray}\label{harml}
  u^{++}_{\M} &\equiv& u^{0}_{\M} + u^{D-1}_{\M} =
    u_{0\M} - u_{D-1\M} = u_{--\M}, \\
  u^{--}_{\M} &\equiv& u^{0}_{\M} - u^{D-1}_{\M} =
    u_{0\M} + u_{D-1\M} = u_{++\M}~. \nonumber
  \end{eqnarray}

The constraints \p{ortho}, \p{compl} as they stand
are invariant under the local $SO(1,D-1)$ transformations acting on
the tangent space indices $\A, \B, ...$.
Below we will see that for the purpose of constructing the geometric
approach description of the string theory this symmetry should be restricted
to $SO(1,1) \otimes SO(D-2)$ in accordance with the splitting \p{harm}.
Just this local symmetry is respected
by the basic geometric postulate of such a description, namely by the
condition that the Lorentz harmonic frame be adapted to the string
worldsheet (see Eq. \p{dx=eu} below). This means that two of
$(D-2)$ vectors $  u^{~\A}_{\M}$
are chosen to be tangent to the worldsheet while the remaining ones
$u^{i}_{\M}$ are orthogonal to it. The local (gauge) symmetry
$SO(1,1) \otimes SO(D-2)$ reflects, respectively,
the freedom of the $d=2$ Lorentz rotation of
 the vectors $u_{\M}^{0,(D-1)}$
tangent to the worldsheet and $SO(D)$ rotations of the vectors $u^i_{\M}$
orthogonal to the worldsheet (in the light-like notation \p{harml}
the $d=2$ rotations are reduced to the opposite weight scaling
transformations of the vectors $u^{++}$ and $u^{--}$).

Thus  the vectors $u^{\A}_{\M}$ appropriate for the description of the
external geometry of the bosonic string worldsheet
parametrize the {\sl non-compact} coset space
  \begin{equation}\label{coset}
  \frac{SO(1,D-1)}{SO(1,1) \times SO(D-2)}~.
  \end{equation}

In other words,
the harmonics \p{harm} regarded as the worldsheet fields
define a map of the worldsheet ${\cal M}^{(1,1)} = \{ \xi^m \}$
onto the non-compact coset
\p{coset}
 \begin{equation}\label{ucoset}
 u^{\A}_{\M}~~~: ~~~ {\cal M}^{(1,1)} = \{ \xi^m \}
~~\rightarrow ~~~\frac{SO(1,D-1)}{SO(1,1) \times SO(D-2)}~.
  \end{equation}

Below we will see that the basic ingredients of the sought general solution
of the string-inspired equations will be smaller sets of left- and right-moving
Lorentz harmonics parametrizing some compact subspaces in the chiral
copies of the coset space \p{coset}.

\subsubsection{Cartan 1--forms}

Differentials of the harmonic variables can be calculated with taking
into account the conditions \p{harm}.
Differentiating Eq. \p{harm} produces  the
equation
$$
 d u^{\underline{a}}_{\underline{m}}
 u^{~\underline{b}\underline{m}}
+
 u^{~\underline{a}\underline{m}}
 d u^{\underline{b}}_{\underline{m}}
= 0
$$
which can be solved as follows
 \begin{equation}\label{hdif}
 d u^{~\underline{a}}_{\underline{m}} =
 u^{~\underline{b}}_{\underline{m}}
 \Om^{~\underline{a}}_{\underline{b}} (d)
 \qquad  \Leftrightarrow \qquad
 \cases {
 du^{~++}_{\underline{m}}
 = u^{~++}_{\underline{m}}  \om
 + u^{~i}_{\underline{m}} f^{++i} (d ) ,  \cr
 du^{~--}_{\underline{m}}
 = - u^{~--}_{\underline{m}}  \om
 + u^{~i}_{\underline{m}} f^{--i} (d ) ,
 \cr
 d u^{i}_{\underline{m}} = - u^{j}_{\underline{m}}  A^{ji} +
 {1\over 2} u_{\underline{m}}^{++} f^{--i} (d) +
 {1\over 2} u_{\underline{m}}^{--} f^{++i} (d)~.
\cr }
\end{equation}
Here

\begin{equation}\label{pC}
 \Om^{~\underline{a}}_{\underline{b}}
  \equiv
u_{\underline{b}}^{\underline{m}} d u^{\underline{a}}_{\underline{m}}
 = \pmatrix {
\om &  0 & {1 \over \sqrt{2}} f^{--i} (d ) \cr
  0 & -\om & {1 \over \sqrt{2}} f^{++i} (d ) \cr
 {1 \over \sqrt{2}} f^{++i} (d ) &
 {1 \over \sqrt{2}} f^{--i} (d ) &
                         A^{ji} (d) \cr} , \qquad
\Om^{\underline{a}\underline{b}}
\equiv
\eta^{~\underline{a}\underline{c}}
\Om^{~\underline{b}}_{\underline{c}} =
- \Om^{\underline{b}\underline{a}}
\end{equation}
are the $SO(1,D-1)$ Cartan forms (in the vector representation of the
$SO(1,D-1)$ generators). Due to \p{harm}, they
are naturally divided into the
$SO(1,1)\times SO(D-2)$ covariant forms
\begin{equation}\label{+i}
f^{++i} \equiv
u^{++}_{\underline m} d u^{\underline{m}i},
\end{equation}

\begin{equation}\label{-i}
 f^{--i} \equiv u^{--}_{\underline m} d u^{\underline{m}i} ,
\end{equation}
which constitute a vielbein of the non-compact coset space
$
{SO(1,D-1) \over SO(1,1) \otimes SO(D-2)}
$,
the $SO(1,1)$ (spin) connection
\begin{equation}\label{0}
\omega \equiv {1 \over 2} u^{--}_{\underline m} d
 u^{\underline  m~++}\; ,
\end{equation}
and the $SO(D-2)$ connections
(gauge fields)
\begin{equation}\label{ij}
 A^{ij} \equiv u^{i}_{\underline m} d u^{\underline{m}~j} \equiv 
- u^{j}_{\underline{m}}  d  u^{\underline{m}i}\; .
\end{equation}

\subsubsection{Parabolic subgroup}

As was mentioned above, the choice of the tangent space local group
as $SO(1,1) \otimes SO(D)$ (and the coset space \p{coset} for
Lorentz harmonics) is motivated by the adaptation postulate
which ``solders'' harmonics to the worldsheet and is so relevant just to the
geometric description of strings.
Formally, the same harmonic (moving frame) variables \p{harm} can be used
to give a geometric description of the massless particle.
But in this case the adaptation of the moving frame
would consist in requiring that one of the light--like vectors, e.g.
$u^{++}_{\M}$, is tangent to the worldline, while $u^i$ and $u^{--}$
are orthogonal to it.
Such an adaptation is covariant with respect to the following right gauge
transformations \cite{gds} (see also \cite{BandH}, where the Hamiltonian form
of the corresponding $D=4$ transformations has been presented)
  \begin{eqnarray}\label{triangR}
  u^{++}_{\M}{}^{\prime} &=& u^{++}_{\M} V^{-1}, \nonumber \\
  u^{i}_{\M}{}^{\prime}  &=&( u^{++}_{\M} V^{--j} +
    u^{j}_{\M}) V^{ji}, \\
  u^{--}_{\M}{}^{\prime} &=& ( u^{--}_{\M}+ u^{++}_{\M} V^{--i}V^{--i} +
    2 u^{i}_{\M}  V^{--j} ) V~. \nonumber
  \end{eqnarray}
The transformations \p{triangR}
form the maximal proper subgroup (parabolic subgroup)
$SO(1,1) \times SO(D-2)
\semiprod K_{D-2}$ of the Lorentz group $SO(1,D-1)$ \cite{gds}.
An arbitrary element of
this subgroup is characterized by
the $SO(1,1)$ transformation $V=e^\a$, the matrix $V^{ij}$ of
$SO(D-2)$-orthogonal rotations and the parameters
$V^{--i}$
of the boosts $K_{D-2}$.
Thus, in the case of geometric description of massless particle,
{\sl as well as in any case when the adaptation of the moving frame
involves only one light--like moving frame vector},
the harmonics
$u^{\A}_{\M}$
\p{harm} can be regarded as parameters
of the {\sl compact}
coset space
  \begin{equation}\label{sphere}
  S^{D-2} = \frac{SO(1,D-1)}{SO(1,1) \times SO(D-2) \semiprod K_{D-2}}~.
  \end{equation}
It is isomorphic to a $(D-2)$--dimensional  sphere $S^{D-2}$ \cite{gds}.

The Cartan forms \p{+i}, \p{-i}, \p{0}, and \p{ij} are transformed
under \p{triangR} as follows
  \begin{equation}\label{f+tr}
  f^{++i\prime} = f^{++j} V^{ji} V^{-1},
  \end{equation}
  \begin{equation}\label{f-tr}
  f^{--i\prime} =
\Big(f^{--j} + 2 {\cal D}V^{--j}
- 2 f^{++k} (V^{--k} V^{--j} -
{1 \over 2} \d^{kj} V^{--l} V^{--l})\Big)V^{ji} ,
\end{equation}
  $$
 {\cal D}V^{--j} \equiv d V^{--j} + \om V^{--j} - V^{--k} A^{kj},
 $$

  \begin{equation}\label{omtr}
  \om^{\prime} = \om - f^{++i} V^{--i} + V dV^{-1},
  \end{equation}
  \begin{equation}\label{aijtr}
  A^{ij\prime} =    (V^{-1} A V)^{ij} - (V^{-1} dV)^{ij}
          -2 f^{++k} V^{k[i|} V^{--l} V^{l|j]}~.
  \end{equation}

\bigskip

Though the $K_{D-2}$ transformations \p{triangR}
with the  parameters $V^{--i}$
{\sl are not } the gauge symmetries of the whole bosonic string theory,
they play a crucial role for understanding the
group-theoretical structure of the general solution of the nonlinear
equations \p{WZNW1}, \p{Liouville0}
(see Section 3)
\footnote{It was noted in \cite{zero} that the boost symmetry allows one to
introduce a non-trivial dependence on a spectral parameter into
the connection 1-forms entering the zero curvature representation
for the nonlinear equations \p{WZNW1}, \p{Liouville0}.}.
Moreover, to define the general solution, we introduce two chiral sets
of moving frame variables, $l^{(\A )}_{\M}$ and $r^{(\A )}_{\M}$
(see Section 2.2). Their ``adaption'' is realized just in the
``particle--like'' fashion and, thus,
respects covariance under the chiral counterparts of the maximal parabolic
symmetry \p{triangR} (cf. \p{tmapR}, \p{tmapL}).

\bigskip

\subsection{The first-order form of string equations
and the geometric \break approach to string theory}

The Lorentz harmonics give us a possibility to rewrite
the string equations of motion in the first order form, namely, as the
following set of equations \cite{bzstr,bpstv,zero}

\begin{equation}
\label{dx=eu}
dX^{\underline{m}} =
{1\over 2} e^{++} u^{--\underline{m}}+ {1\over 2} e^{--} u^{++\underline{m}},
\end{equation}
\begin{equation}
\label{strLh2}
d(e^{++} u^{--\underline{m}} -
e^{--} u^{++\underline{m}}) = 0~.
\end{equation}
Here $e^{\pm\pm} = d\xi^m~e_m^{\pm\pm}$ is a worldsheet vielbein. While
dealing with \p{dx=eu} and \p{strLh2}, one should take into account
the restrictions \p{hdif} on the differentials of the harmonic
variables.

Eq. \p{dx=eu} is the adaptation relation already referred
to in the consideration above. It plays the basic role in the geometric
approach to strings. In particular, it implies that the intrinsic worldsheet
metric is identified with the induced one
  \begin{equation}\label{eform}
  \frac{1}{2}(e^{++}_{m} e^{--}_{n} + e^{--}_{m} e^{++}_{n}) =
    g_{mn} = \partial_{m} X^{\M} \partial_{n} X_{\M}~.
  \end{equation}

\bigskip
The geometric meaning of Eq. \p{dx=eu} consists in the statement
that the string worldsheet is a surface embedded into the
$D$--dimensional Minkowski space-time. On its own right,
it has no any dynamical content and gives rise to purely
geometric consequences.

The integrability conditions
($ddX\equiv 0$) for Eq. \p{dx=eu} are as follows
  \begin{equation}\label{tor}
  T^{++} \equiv de^{++} + e^{++} \wedge \om = 0~, \qquad
  T^{--} \equiv de^{--} - e^{--} \wedge \om = 0~,
  \end{equation}
  \begin{equation}\label{sff}
  e^{++} \wedge f^{--i} + e^{--} \wedge f^{++i} = 0 \qquad
    \Leftrightarrow \qquad f^{~--i}_{--} - f^{~++i}_{++} = 0~.
  \end{equation}
Thus they require the torsion  2--forms  $T^{\pm\pm}$ to vanish, thereby
imposing proper constraints on the induced spin connection $\om $.
Besides, they imply the $SO(1,1)$ invariant components of the
covariant one--forms $f^{++i}$ and $f^{--i}$ to coincide with each other
 \begin{equation}\label{hi}
 f^{~--i}_{--} = f^{~++i}_{++} = {1 \over 2} h^i~.
  \end{equation}
The quantity $h^i$ can be easily recognized as the main curvatures
of the embedded surface \cite{Ei}.
Indeed, using Eq. \p{dx=eu} to express the light--like harmonic vectors
in terms of derivatives of the embedding functions $X$, we arrive at
the standard expression for the main curvatures
  \begin{equation}\label{curv}
  h^{i} = g^{mn} K^{i}_{mn}, \qquad
    K^{i}_{mn} = - \partial_{m}\partial_{n} X^{\M} u^{i}_{\M}~.
  \end{equation}

\subsubsection{Minimal embedding}

Using Eq. \p{hdif}, one finds that Eq. \p{strLh2} implies
the vanishing of the main curvatures \p{hi}
  \begin{equation}\label{h=0}
  h^{i} = 0~.
  \end{equation}
Thus the surface defined by Eq.
\p{strLh2}
is minimal \cite{Ei}.

On the other hand, expressing all the
auxiliary variables $u$ and $e$ through the derivatives
of the embedding functions $X(\xi )$ and
using the induced metric \p{eform}
\begin{equation}\label{indm0}
g_{mn} \equiv
\partial_m X^{\underline{m}} \partial_n X_{\underline{m}}~,
\end{equation}
and its inverse $g^{mn}$, one can rewrite
Eq. \p{h=0} (or \p{strLh2}) in the form
\begin{equation}\label{str}
\partial_m \big( \sqrt{-g} g^{mn} \partial_n X^{\underline{m}} \Big) = 0~,
\end{equation}
which is the standard string equations of motion pertinent to the
Nambu--Goto action (see \cite{zero} for details).
Thus the strings dynamics in the geometric approach is contained
just in eq. \p{strLh2}.

\bigskip

Note that it is important for the geometric approach description
that the minimal embedding is
described by the covariant Cartan forms
$f^{++i}$, $f^{--i}$ containing in their decomposition only one of
the two basic forms $e^{--}$,  $e^{++}$
  \begin{equation}\label{fpmemp}
  f^{++i} =  e^{--} f_{--}^{~++i}~, \qquad
  f^{--i} =  e^{++} f_{++}^{~--i}
  \end{equation}
(cf. Eqs. \p{hi}, \p{h=0}). Actually, Eqs. \p{fpmemp} encode three
previous equations: \p{sff}, \p{hi} and \p{h=0}. Thus 
the string dynamics is finally encoded just in Eqs. \p{fpmemp}.

\subsection{Maurer-Cartan equation and the string-inspired nonlinear
\break equations}

The integrability conditions for Eqs. \p{hdif} produce
the Maurer-Cartan equations
  \begin{equation}\label{MC}
  d\Om^{\A\B} - \Om^{\A}_{~\C} \wedge \Om^{\C\B} = 0~.
  \end{equation}
Owing to Eq. \p{harm}, Eqs. \p{MC} naturally split into the
following equations for the coset vielbeins
$f^{\pm\pm i}$ \p{+i}, \p{-i}
and the connection one-forms $\om $, $A^{ij}$
\p{0}, \p{ij}
  \begin{equation}\label{PC+}
  {\cal D} f^{++i} \equiv d f^{++i} - f^{++i} \wedge \om +
    f^{++j} \wedge A^{ji} = 0~,
  \end{equation}
  \begin{equation}\label{PC-}
  {\cal D} f^{--i} \equiv d f^{--i} + f^{--i} \wedge \om +
    f^{--j} \wedge A^{ji} = 0~,
  \end{equation}
  \begin{equation}\label{G}
    {\cal R} \equiv d \om = {1\over 2} f^{--i} \wedge f^{++i}~,
  \end{equation}
  \begin{equation}\label{R}
  R^{ij} \equiv d A^{ij} + A^{ik} \wedge A^{kj} =
    - f^{--[i} \wedge f^{++j]}~.
  \end{equation}
Eqs. \p{PC+} --\p{R} amount to the Peterson-Codazzi, Gauss
and Ricci equations of the classical surface theory \cite{Ei}.

Thus, in the geometric approach framework, the dynamics
of string is described by the vielbein $e^{\pm\pm}$ and the set of Cartan
forms $\om$, $f^{\pm\pm i}$, $A^{ij}$ which satisfy
Eqs. \p{tor}, \p{fpmemp} and the Maurer-Cartan
Eqs. \p{PC+} -- \p{R} \cite{Ei,geom,BNbook,bpstv,zero}.

The most essential general feature of this approach is that Eqs.
\p{fpmemp}, \p{tor}, \p{PC+}, \p{PC-} can be solved
algebraically \cite{zero}.

Indeed, using the fact that any connection is integrable on any
one-dimensional subspace, we can specify the expressions for the
$SO(1,1)$ and $SO(D-2)$ connection one-forms in the following way
  \begin{equation}\label{omcov}
  \om =  e^{++} \nabla_{++} (W-L) -
         e^{--} \nabla_{--} (W+L)~,
  \end{equation}
  \begin{equation}\label{aijcov}
  A^{ij} = e^{++} \nabla_{++} \GR^{ik} \GR^{jk} +
           e^{--} \nabla_{--} \GL^{ik} \GL^{jk}~,
  \end{equation}
where $\GL$ and $\GR$ are some $SO(D-2)$ group  matrices,
  \begin{equation}\label{GGT}
  \GL \GL^T = I, \qquad \GR \GR^T = I~.
  \end{equation}
Eqs. \p{omcov}, \p{aijcov} provide a possibility to rewrite
Eqs. \p{tor}, \p{PC+}, \p{PC-} as
the conditions of closeness of some one-forms
  \begin{equation}\label{tor1}
  d(e^{++} \exp(W+L)) = 0~, \qquad
  d(e^{--} \exp(W-L)) = 0~.
  \end{equation}
  \begin{equation}\label{PCclos}
  d\left(f^{++i} \GR^{ji} \exp(-W+L)\right) = 0~, \qquad
  d\left(f^{--i} \GL^{ji} \exp(-W-L)\right) = 0
  \end{equation}
(Eqs. \p{fpmemp} have been used when deriving \p{PCclos}).

The general solution of Eqs. \p{tor1}
(up to some possible topological subtleties which are
unessential for the present study) is provided by
  \begin{equation}\label{edxi}
  e^{++} = d\kp \Mpp(\kp) \exp(-W-L)~, \qquad
  e^{--} = d\km \Mmm(\km) \exp(-W+L)~.
  \end{equation}
Here  $\kpm$ are some functions of the string worldsheet
coordinates with the only defining demand that their
differentials are linearly-independent one-forms.
It is convenient, however, to choose $\kpm$ as a set of
local coordinates on the worldsheet
  \begin{equation}\label{xigauge}
  \xi^{m} = \xi^{(\pm\pm)}~.
  \end{equation}
This choice fixes the gauge with respect to the worldsheet
reparametrizations (general coordinate
transformations) so that only 2-dimensional conformal
reparametrizations survive (see Eq. \p{conf1} below).

In the holonomic
basis $d\kpm$ the components of the vielbein form $e^{\pm\pm}$ are given by
  \begin{equation}\label{ecomp}
  \begin{array}{ll}
  e^{++}_{(++)} = \Mpp \exp(-W-L)~, & e^{++}_{(--)} = 0~, \\
  e^{--}_{(++)} = 0~, & e^{--}_{(--)} = \Mmm \exp(-W+L)~, \\
  e^{(++)}_{--} = 0~, & e^{(--)}_{--} = 2(\Mmm)^{-1} \exp(W-L)~, \\
  e^{(++)}_{++} = 2(\Mpp)^{-1} \exp(W+L)~, & e^{(--)}_{++} = 0~,
  \end{array}
  \end{equation}
the induced metric \p{eform} is conformally flat
  \begin{equation}\label{metricc}
  ds^2 = d\xi^m d\xi^n g_{mn}
= d\xi^{(++)} d\xi^{(--)} \Mpp \Mmm e^{2W}
  \end{equation}
and the covariant derivatives are proportional to the corresponding
holonomic ones
  \begin{equation}\label{dder}
  e^{++} \nabla_{++} =
    d \kp \partial_{(++)}, \qquad
 e^{--} \nabla_{--} =
    d \km \partial_{(--)}~.
  \end{equation}
Hence,  the expressions \p{omcov} and \p{aijcov} for the gauge connections
can be rewritten as follows
  \begin{equation}\label{omhol}
  \om = d\kp \pap (W-L) - d\km \pam (W+L)~,
  \end{equation}
  \begin{equation}\label{aijhol}
  A^{ij} = d\kp \pap \GR^{ik} \GR^{jk} + d\km \pam \GL^{ik} \GL^{jk}~.
  \end{equation}

Finally, using the holonomic basis, one  can write down the general solution
of the Peterson--Codazzi equations \p{PCclos} for the
covariant forms \p{fpmemp} as
  \begin{equation}\label{fdxi+}
  f^{++i} = d\km e^{W-L} \GR^{ij} M^{(++)j}_{(--)} (\km)~,
  \end{equation}
  \begin{equation}\label{fdxi-}
  f^{--i} = d\kp e^{W+L} \GL^{ij} M^{(--)j}_{(++)}(\kp)~,
  \end{equation}
where vector fields $M^{(++)j}_{(--)}$ , $M^{(--)j}_{(++)}$
are chiral, similarly to the parameters $\Mpp$, $\Mmm$ of the
solutions \p{edxi} of Eqs. \p{tor}
$$
\partial_{(++)}M^{(++)j}_{(--)} = 0~, \qquad
\partial_{(--)}M^{(--)j}_{(++)} = 0~. 
$$

\bigskip

Thus, following \cite{zero}, we have solved algebraically all the
equations except for the Gauss and Ricci ones \p{G}, \p{R}.
Substituting \p{omhol}, \p{aijhol}, \p{fdxi+}, \p{fdxi-}
into Eqs. \p{G} and \p{R}, we obtain the set of nonlinear equations
  \begin{equation}\label{Liouville}
  \pap \pam W =
    {1 \over 4} M^{(++)i}_{(--)} (\GL^T\GR)^{ij} M^{(--)j}_{(++)} e^{2W}~,
  \end{equation}
  $$
  \pam \left((\pap \GL) \GL^T \right)^{ij} -
  \pap \left((\pam \GR) \GR^T \right)^{ij} +
  \left[(\pap \GL) \GL^T ~,~ (\pam \GR) \GR^T \right]^{ij} =
  $$
  \begin{equation}\label{Smodel}
  = e^{2W} (\GL M_{(--)}^{(++)})^{[i} (\GR M_{(++)}^{(--)})^{j]}
  \end{equation}
describing the extrinsic geometry of the string worldsheet embedded into
a D--dimensional Minkowski space.

\subsection{Zero curvature representation and the associated linear system}

As we saw, most of the equations of the geometric approach
\p{fpmemp}, \p{tor}, \p{PC+}, \p{PC-} can be solved algebraically
and the final set of the string-inspired nonlinear equations
\p{Liouville}, \p{Smodel} emerges as the result of substitution of
these algebraic solutions into the Gauss and Ricci equations
\p{G}, \p{R}. Since the latter constitute a part of the Maurer--Cartan
equation \p{MC}, one can conclude that the zero curvature
representation for the nonlinear equations \p{Liouville}, \p{Smodel}
is given by the Maurer--Cartan equation \p{MC} for the $so(1,D-1)$
valued connection one-forms specified
by Eqs. \p{pC}, \p{omhol}, \p{aijhol}, \p{fdxi+}, \p{fdxi-}.

The associated linear system is provided by Eq. \p{hdif} with the
forms \p{MC} specified
by Eqs.
\p{pC}, \p{omhol}, \p{aijhol}, \p{fdxi+}, \p{fdxi-}.

A non-trivial dependence on a spectral parameter can be introduced
into the associated linear system by means of the
parabolic subgroup transformations \p{triangR} \cite{zero}.

Thus the nonlinear equations \p{Liouville}, \p{Smodel}
possess all the features inherent to the equations which can
be solved by the Inverse Scattering Method
\cite{Faddeev}.
Below we will prove that they are solvable even in a more
strong sense, like the Liouville, Toda or WZNW sigma-model equations.
Namely, we will deduce an explicit form of
the general solution. It is interesting that the solution can be obtained
by exploiting the parabolic group transformations \p{triangR}. We
will demonstrate this in the last section of the present paper.

\subsection{Symmetries and bridges}

The obtained nonlinear equations and their zero curvature representation
\p{MC}, \p{pC}, \p{omhol}, \p{aijhol}, \p{fdxi+}, \p{fdxi-} possess
a number of powerful symmetries.

As was already mentioned, the local (gauge) symmetries of the string
model form the $SO(1,1) \times SO(D-2)$ group
  \begin{equation}\label{omtr1}
  \om^{\prime} = \om + V dV^{-1}= \om + \a~ ,
  \end{equation}
  \begin{equation}\label{aijtr1}
  A^{ij\prime} = (V^{-1} A V)^{ij} - (V^{-1} dV)^{ij}~,
  \end{equation}
  \begin{equation}\label{f+trhom}
  f^{++i\prime} = V^{-1} f^{++j} V^{ji}~,
  \end{equation}
  \begin{equation}\label{f-trhom}
  f^{--i\prime} = V f^{--j} V^{ji}~.
  \end{equation}
The matrix fields $\GLR$ appearing in
Eqs. \p{fdxi+}, \p{fdxi-} and \p{aijhol} are transformed homogeneously
under the $SO(D-2)$ gauge symmetry
  \begin{equation}\label{GLRtr}
  \GLR^{ij\prime} = \GLR^{kj} V^{ki} = (V^{-1})^{ik} \GLR^{kj}~,
  \end{equation}
whereas the field $L$ is pure gauge
  \begin{equation}\label{Ltr}
  L^{\prime} = L - \a~, \qquad V = e^{-\a}~.
  \end{equation}
In other words, it is a compensator (or Nambu-Goldstone field)
for the $SO(1,1)$ gauge symmetry.

Examining the expressions \p{edxi},
\p{fdxi+}, and \p{fdxi-}, one concludes that our system possesses
two types of the
infinite-dimensional global
symmetries whose parameters
can be combined into left-moving and right-moving (chiral) functions.
One of them is the $d=2$ {\sl conformal symmetry}
  \begin{equation}\label{conf1}
  d\kp{}^\prime = d\kp s_L(\kp)~, \qquad
  d\km{}^\prime = d\km s_R(\km)~,
  \end{equation}
  $$
  \pam s_L = 0, \qquad \pap s_R = 0~. \qquad
  $$
The second one is realized as chiral rescalings of the chiral fields
$\Mpp$ and $\Mmm$ present in Eqs. \p{fdxi+}, \p{fdxi-}
  \begin{equation}\label{conf2}
  \Mpp{}^\prime = \Mpp e^{h_L}~, \qquad
  \Mmm{}^\prime = \Mmm e^{h_R}~,
  \end{equation}
  $$
  \pam h_L = 0~, \qquad \pap h_R = 0~.
  $$
Clearly, these rescalings can be treated as a sort of affine (or Kac-Moody)
$SO(1,1)_L$ and $SO(1,1)_R$ symmetry transformations (i.e. as $SO(1,1)$
transformations with the parameters depending on $\xi^{(++)}$ and
$\xi^{(--)}$, respectively).

The full set of non-trivial transformations of these symmetries on
the involved fields is given by
  \begin{equation}\label{conftr}
  \Mpp{}^\prime = \Mpp s_L^{-1} e^{h_L}~, \qquad
  \Mmm{}^\prime = \Mmm s_R^{-1} e^{h_R}~,
  \end{equation}
  $$
  (W+L)^\prime = (W+L) + h_L -\alpha~, \qquad
  (W-L)^\prime = (W-L) + h_R + \alpha~,
  $$ $$
  (M_{(++)}^{(--)i})^\prime = M_{(++)}^{(--)i} s_L^{-1} e^{-h_L}~, \qquad
  ( M_{(--)}^{(++)i})^\prime = M_{(--)}^{(++)i} s_R^{-1} e^{-h_R}~.
  $$
We observe that the chiral fields $\Mpp$~, $\Mmm$ can be regarded as
the ``bridges'' relating the affine  $SO(1,1)_L$, $SO(1,1)_R$ groups to
the corresponding chiral parts of the 2-dimensional conformal group,
while
$$e^{W+L} = (e^{W+L})_{++}^{(++)}~, \qquad e^{W-L} = (e^{W-L})_{--}^{(--)}
$$
as the bridges relating affine $SO(1,1)_L$ and $SO(1,1)_R$ to the
gauge $SO(1,1)$ symmetry.
Since the symmetries \p{conf1}, \p{conf2} offer the possibility to choose
the gauge 
\footnote{They also make it possible to fix the value of the norm of the
chiral vector fields $\Mpmmp{i}$ (see \cite{zero}).} $\Mpp = \Mmm = 1$ ,
in what follows, for simplicity, we will make no distinction between the
$SO(1,1)_{L,R}$ indices and conformal ones.
We will also use the superscript $(--)$ instead of
the $SO(1,1)_L$ subscript $(++)$  for the chiral vector field
$M^{(--)i}_{(++)}$ (the chirality of the latter field,
$\partial_{(--)}M^{(--)i}_{(++)}=0$, excludes any confusion).

In addition, our equations possess an
invariance under right multiplication of the $SO(D-2)$
valued fields $\GL$ and $\GR$ by {\sl chiral} $SO(D-2)$ matrices
$H_R$ and $H_L$. So, the complete form of the appropriate symmetry
transformations is
  \begin{equation}\label{SO3}
  \GL{}^{\prime} = V^{-1} \GL H_L~, \qquad
  \GR{}^{\prime} = V^{-1} \GR H_R~,
  \end{equation}
  \begin{equation}\label{transfM}
  M_{(++)}^{(--)i~\prime} = H_L^{ij} M_{(--)}^{(++)j}~, \qquad
  M_{(++)}^{(--)j}{}^\prime = H_R^{ij} M_{(++)}^{(--)j}~,
  \end{equation}
  \begin{equation}\label{Hfield}
  H_L H_L^T = H_R H_R^T = I~, \qquad
  \pam H_L = \pap H_R = 0~.
  \end{equation}
Thus the orthogonal matrix fields $\GL$ and $\GR$ can be regarded as
bridges between the gauge $SO(D-2)$ transformations and affine
chiral $SO(D-2)_L$ and  $SO(D-2)_R$ transformations, respectively.

\subsection{Simplified form of the nonlinear equations}

The system of nonlinear equations \p{Liouville}, \p{Smodel}
can be significantly simplified by using
the $SO(D-2)$ gauge symmetry with parameters $V^{ij}$ to
fix a gauge
  \begin{equation}\label{gauge}
  \GL = 1~, \qquad \GR = G~.
  \end{equation}

Then the  sigma-model-type equation \p{Smodel} acquires the
simplest WZNW sigma-model-type form \p{WZNW1}
  \begin{equation}\label{WZSmodel}
  \pam \left((\pap G) G^T\right)^{ij} =
    e^{2W} G^{[i|k} M_{(--)}^{(++)k}
M_{(++)}^{(--)|j]}~,
  \end{equation}
 whereas the Liouville-type equation \p{Liouville} becomes \p{Liouville0}
  \begin{equation}\label{Liouville1}
  \pap \pam W =
    {1 \over 4} M_{(++)}^{(--)i}  G^{ij} M_{(--)}^{(++)j}  e^{2W}~.
  \end{equation}

The gauge \p{gauge} is invariant under the action of
two chiral affine $SO(D-2)$ symmetries with the parameters $H_L(\kp)$
and $H_R(\km)$, action of which on the matrix field $G$ is given by
  \begin{equation}\label{SO2}
  G^{\prime} = H_L^{-1} G H_R~, \qquad
  \pam H_L = \pap H_R = 0~.
  \end{equation}

\section{General solution of the string-inspired nonlinear \break equations}

\subsection{Standard string equations of motion, their solution and
Virasoro constraints}

We start by discussing the familiar string equations of motion
and Virasoro conditions in the standard setting.
The study of the relationship of the solutions of these equation
with the Lorentz harmonics (which, as was mentioned above, provide
the associated linear system for the considered nonlinear equations)
opens a possibility to construct the general solution of the
nonlinear equations \p{Liouville1}, \p{WZSmodel}.

The equations of motion of $D$-dimensional bosonic string following
from the Nambu-Goto action (see \cite{gsw} and Refs. therein) has the
form \p{str}
  \begin{equation}\label{str0}
  \partial_m \left( \sqrt{-g} g^{mn} \partial_n X^{\M} \right) = 0~,
  \end{equation}
where (recall  eq. \p{indm0})
  \begin{equation}\label{indm}
  g_{mn} \equiv \partial_m X^{\M} \partial_n X_{\M}
  \end{equation}
is the induced metric,  $g^{mn}$ is its
inverse  and  $g  = \det (g_{mn})$.  

In the conformal gauge (see, e.g. \cite{gsw})
the string equation \p{str0} becomes linear
  \begin{equation}\label{str01}
  \tilde{\partial}_{(++)}  \tilde{\partial}_{(--)} X^{\M} \equiv
    {\partial \over \partial  \tilde{\xi}^{(++)}}
    {\partial \over \partial  \tilde{\xi}^{(--)}} X^{\M} = 0
  \end{equation}
and has a general solution
  \begin{equation}\label{strsol}
  X^{\M} = X^{\M}_L + X^{\M}_R~, \qquad
  \tilde{\partial}_{(--)} X^{\M}_L = 0~, \qquad
\tilde{\partial}_{(++)} X^{\M}_R = 0~.
  \end{equation}
The chiral functions $X^{\M}_L ( \tilde{\kp}), X^{\M}_R ( \tilde{\km})$
are subjected to the Virasoro constraints
  \begin{equation}\label{Virasoro}
  \tilde{\partial}_{(++)} X_L^{\M} \tilde{\partial}_{(++)} X_{L\M} = 0~, \qquad
  \tilde{\partial}_{(--)} X_R^{\M} \tilde{\partial}_{(--)} X_{R\M} = 0~.
  \end{equation}

Let us compare the solution \p{strsol} with the expressions \p{dx=eu} and
\p{edxi} obtained in the geometric approach
  \begin{equation}\label{strLh12}
  dX^{\M} = {1\over 2} d\kp e^{-W-L} \Mpp u^{--\M} +
            {1\over 2} d\km e^{-W+L} \Mmm u^{++\M}~.
  \end{equation}

Since, due to Eq. \p{edxi}, the induced metric is conformally flat in the
coordinate frame $\xi^{(\pm\pm)}$,
we can identify these worldsheet coordinates with the ``conformal
coordinates'' $\tilde{\xi}^{(\pm\pm)}$ used in the standard string
description \p{str0}--\p{Virasoro}
$$ \xi^{(\pm\pm)} = \tilde{\xi}^{(\pm\pm)}~.$$
Thus
Eqs. \p{strLh12} and \p{strsol} result in
  \begin{equation}\label{sol-+}
  \pap X_L^{\M}(\kp) =
    {1 \over 2} e^{-W-L} \Mpp(\kp) u^{--\M}~,
  \end{equation}
  $$
  \pam X_R^{\M}(\km) = {1\over 2} e^{-W+L} \Mmm(\km) u^{++\M}~.
  $$
It is easy to verify that the Virasoro constraints \p{Virasoro} are
satisfied for the functions \p{sol-+}.

It is worth noticing that the l.h.s. of Eqs. \p{sol-+} includes the
chiral functions only (cf. \p{strsol}),
while the r.h.s. involve the functions $W$, $L$, $u^{\pm\pm }$ which
from the very beginning were assumed to depend on both coordinates
$\kpm$.

We will demonstrate below that the origin of this fact lies in that
any solution of the string equation produces a
solution of the string--inspired nonlinear equations, i.e. the equations
\p{Liouville}, \p{WZSmodel} which describe the extrinsic geometry
of the string worldsheet.

\subsection{Chiral harmonics}
In ref. \cite{Kapustn} it was discussed how to find
an appropriate set of chiral functions for obtaining an explicit expression
for the $W$ and $G^{ij} = (G^{-1})^{ji}$ fields which enter the nonlinear
equations \p{Liouville}, \p{WZSmodel}. Constrained twistors have been
proposed as such variables for the case of bosonic string theories in
dimensions $D=3,4,6$. Such twistors can be
regarded as spinor Lorentz harmonics \cite{gds,bzstr}. Their only
property to be essential for our purposes is
that they can be used to define the appropriate vector moving
frame. This allows one to avoid complicated
calculations associated with the use of the spinor moving frame or
spinor harmonic formalism \cite{bzstr} and to obtain the solution of
nonlinear equations describing the extrinsic geometry of bosonic string
moving in the Minkowski space of {\sl any} dimension $D$ in terms of the
vector Lorentz harmonics only.

\bigskip

Let us introduce two extra sets of Lorentz harmonics (moving
frame variables)
  \begin{equation}\label{chiharm}
  r^{(\A )}_{\M} (\km ) = \left( r^{(\pm\pm )}_{\M}, r^{i}_{\M} \right)
  \quad \in \quad SO(1,D-1)~,
  \end{equation}
  $$
  l^{(\A )}_{\M} (\kp ) = \left( l^{(\pm\pm )}_{\M}, l^{i}_{\M} \right)
  \quad \in \quad SO(1,D-1)~,
  $$
each depending only on the $\km$ and $\kp$ coordinates of the string
worldsheet.

Recall that the condition \p{chiharm} means
  \begin{equation}\label{orthoR}
  r^{\M}_{~(\A)} r_{\M\B} = \eta_{(\A)(\B)} \equiv \mbox{diag} (1,-1,...,-1)~,
  \end{equation}
$$
~~~~~~~~~~~~~~~~~~~~~~~~\Leftrightarrow \cases{
r_{\M}^{(++)}  r^{\M (++)} = 0~, ~~~  r_{\M}^{(--)}  r^{\M (--)} = 0~, ~~~\cr
r_{\M}^{(++)}  r^{\M (--)} = 2~, ~~~ \cr
r_{\M}^{(++)}  r^{\M i} = 0~, ~~~~ r_{\M}^{(--)}  r^{\M i} = 0~, ~~~\cr
r_{\M}^{i}  r^{\M j} = - \d^{ij}~. ~~~\cr
}
$$
  \begin{equation}\label{orthoL}
  l^{\M}_{~(\A)} l_{\M\B} = \eta_{(\A)(\B)} \equiv \mbox{diag} (1,-1,...,-1)~,
  \end{equation}
$$
~~~~~~~~~~~~~~~~~~~~~~~~\Leftrightarrow \cases{
l_{\M}^{(++)}  l^{\M (++)} = 0~, ~~~  l_{\M}^{(--)}  l^{\M (--)} = 0~, ~~~\cr
l_{\M}^{(++)}  l^{\M (--)} = 2~, ~~~ \cr
l_{\M}^{(++)}  l^{\M i} = 0~, ~~~~ l_{\M}^{(--)}  l^{\M i} = 0~, ~~~\cr
l_{\M}^{i}  l^{\M j} = - \d^{ij}~. ~~~\cr
}
$$
Further, we identify the chiral vectors $\pam X_R^{\M}$ and $\pap X_L^{\M}$
with the components $r^{(++)} = r^{(++)}(\km)$ and
$l^{(--)} = l^{(--)}(\kp)$
of these sets of chiral harmonics
  \begin{equation}\label{dXrl}
  \pap X_L^{\M} = {1 \over 2} \Mpp(\kp) l^{(--)\M}(\kp)~, \qquad
  \pam X_R^{\M} = {1 \over 2} \Mmm(\km) r^{(++)\M}(\km)~.
  \end{equation}
In such a way we adapt the chiral frames to the left and
right sectors  of the string worldsheet.
Since other components of the left- and right-moving frame variables
$l^{(\A )}_{\M }$, $r^{(\A )}_{\M }$ remain
arbitrary, we have just a ``particle--like'' situation in the present case.
Hence,  Eqs. \p{dXrl} possess the invariance under the 
affine
$$(SO(1,1) \otimes SO(D-2) \semiprod K_{D-2})_L \qquad
\mbox{and} \qquad (SO(1,1) \otimes SO(D-2) \semiprod K_{D-2})_L$$
symmetries with chiral parameters
$$V_{L}(\xi^{(++)})= e^{h_L}, \qquad V^{(++)i}_{R}(\xi^{(++)}), \qquad
V^{ij}(\xi^{(++)})=V^{-1ji}(\xi^{(++)})$$
and
$$V_{R}(\xi^{(--)})=e^{h_R}, \qquad  V^{(--)i}_{R}(\xi^{(--)}), \qquad
V^{ij}(\xi^{(--)})=V^{-1ji}(\xi^{(--)})$$
respectively\footnote{
The affine symmetry $SO(1,1)_L \otimes SO(1,1)_R$ proves to be
``soldered'' to the worldsheet conformal symmetry
when the gauge $M^{~(++)}_{(++)}= 1 =M^{~(--)}_{(--)}$ is imposed.}. 
The corresponding transformations read

 \begin{eqnarray}\label{tmapR}
  r^{(++)\prime }_{\M} &=& r^{(++)}_{\M} V_{R}~,
\nonumber \\
  r^{i\prime}_{\M}  &=& (r^{j}_{\M} + r^{(++)}_{\M} V^{(--)i}_{R}
) V^{ji}_{R}~,
 \\
  r^{(--)\prime}_{\M} &=&
   ( r^{(--)}_{\M} +  r^{(++)}_{\M} V^{(--)i}_{R} V^{(--)i}_{R} +
2  r^{i}_{\M} V^{--i}_{R}) V^{-1}_{R}~, \nonumber
  \end{eqnarray}
  \begin{eqnarray}\label{tmapL}
  l^{(--)}_{\M} &=& l^{(--)}_{\M} V_{L}~, \nonumber \\
  l^{i}_{\M}  &=&
( l^{j}_{\M}  + l^{(--)}_{\M} V^{(++)i}_{L}
)V^{ji}_{L}~, \\
  l^{(++)}_{\M} &=&
     ( l^{(++)}_{\M} + l^{(--)}_{\M} V^{(++)i}_{L}V^{(++)i}_{L}
+ l^{i}_{\M} V^{(++)j}_{L}  )V^{-1}_{L}~. \nonumber
  \end{eqnarray}

Hence, each set of chiral harmonics parametrizes the sphere \p{sphere}.
As they depend only on one of the worldsheet coordinates,
$\xi^{(--)}$ or $\xi^{(++)}$, they map one of the
light-like sectors, ${\cal M}^{(0,1)} = \{ \xi^{(--)} \}$ or
 ${\cal M}^{(1,0)} = \{ \xi^{(++)} \}$, of the worldsheet
${\cal M}^{(1,1)} = \{ \xi^m \}= \{ \xi^{(++)}, \xi^{(--)} \}$ onto
two copies of this sphere
 \begin{equation}\label{rsphere}
r_{\M}^{(\A)} : \qquad {\cal M}^{(0,1)} = \{ \xi^{(--)} \} ~~ \rightarrow ~~
 S^{D-2} = \frac{SO(1,D-1)}{SO(1,1) \times SO(D-2) \semiprod K_{D-2}}~,
  \end{equation}
  \begin{equation}\label{lsphere}
l_{\M}^{(\A)} : \qquad {\cal M}^{(1,0)} = \{ \xi^{(++)} \} ~~ \rightarrow ~~
 S^{D-2} = \frac{SO(1,D-1)}{SO(1,1) \times SO(D-2) \semiprod K_{D-2}}~.
  \end{equation}
Below we denote the spaces of all possible images of these maps by
$S^{(D-2)}_L$ and $S^{(D-2)}_R$, respectively.

\bigskip

The chiral counterparts of the Cartan forms \p{pC}
contain only one of  the chiral holonomic basic 1--forms $d\xi^{(--)}$
or $d\xi^{(++)}$
  \begin{equation}\label{f++R}
 f^{(++)i}_R = d\xi^{(--)} f^{~(++)i}_{(--)R} = r^{(++)\M}dr^i_{\M}~ ,  \qquad
f^{~(++)i}_{(--)R} = r^{(++)\M}\partial_{(--)} r^i_{\M}
  \end{equation}
  \begin{equation}\label{f--R}
 f^{(--)i}_R = d\xi^{(--)} f^{~(--)i}_{(--)R} = r^{(--)\M}dr^i_{\M}~,   \qquad
f^{~(--)i}_{(--)R} = r^{(--)\M}\partial_{(--)} r^i_{\M}
  \end{equation}
 \begin{equation}\label{omR}
 \om_R = d\xi^{(--)} \om_{(--)R} = {1\over 2}r^{(--)\M}dr^{(++)}_{\M}~,
\qquad \om_{(--)R} = {1\over 2}r^{(--)\M}\partial_{(--)} r^{(++)}_{\M}
  \end{equation}
  \begin{equation}\label{AijR}
 A^{ij}_R = d\xi^{(--)} A^{ij}_{(--)R}
= {1\over 2}r^{i\M}d r^{j}_{\M}~, \qquad
A^{ij}_{(--)R} = {1\over 2}r^{i\M}\partial_{(--)} r^{j}_{\M}
  \end{equation}

\bigskip

  \begin{equation}\label{f++L}
 f^{(++)i}_L = d\xi^{(++)} f^{~(++)i}_{(++)L} = l^{(++)\M}dl^i_{\M}~,   \qquad
f^{~(++)i}_{(++)L} = l^{(++)\M}\partial_{(++)}  l^i_{\M}
 \end{equation}
  \begin{equation}\label{f--L}
 f^{(--)i}_L = d\xi^{(++)} f^{~(--)i}_{(++)L} = l^{(--)\M}dl^i_{\M}~,  \qquad
f^{~(--)i}_{(++)L} = l^{(--)\M}\partial_{(++)}l^i_{\M}
  \end{equation}
  \begin{equation}\label{omL}
 \om_L = d\xi^{(++)} \om_{(++)L} = {1\over 2}l^{(--)\M}dl^{(++)}_{\M}~,  \qquad
\om_{(++)L} = {1\over 2}l^{(--)\M}\partial_{(++)}l^{(++)}_{\M}
  \end{equation}
  \begin{equation}\label{AijL}
 A^{ij}_L = d\xi^{(++)} A^{ij}_{(++)L} = {1\over 2}l^{i\M}dl^{j}_{\M}~,  \qquad
A^{ij}_{(++)L} = {1\over 2}l^{i\M}\partial_{(++)}l^{j}_{\M}~.
  \end{equation}

\bigskip

The transformations of these 1--forms under the left and right
affine $SO(1,1)\otimes SO(D-2) \semiprod K_{D-2}$ symmetry
\p{tmapR} and \p{tmapL}
are determined
by the chiral version of Eqs. \p{f+tr}--\p{aijtr} and its
evident ``left'' counterpart respectively.
It is worth noting that only the forms
$$f^{(++)i}_R = d\xi^{(--)} f^{~(++)i}_{(--)R} \qquad
\mbox{and} \qquad f^{(--)i}_R = d\xi^{(++)} f^{~(--)i}_{(++)L}$$
transform covariantly under \p{tmapR} and \p{tmapL}.
These forms are vielbeins of the 'chiral spheres'
$S^{D-2}_R$ \p{rsphere}
and $S^{D-2}_R$  \p{lsphere}, respectively.

\subsection{Relation of general and chiral harmonics:
solving the \break Liouville--type equation}

Substituting Eq. \p{dXrl} into Eq.\p{sol-+}, one obtains the expression for
the chiral light--like moving frame vector fields $l^{(--)\M} (\kp)$,
$r^{(++)\M} (\km)$
in terms of generic light--like  harmonics $u^{--\M}, u^{++\M}$,
the Liouville field $W$ and compensator $L$
  \begin{equation}\label{rl=u}
  l^{(--)\M} (\kp) = e^{-W-L} u^{--\M}~, \qquad
  r^{(++)\M} (\km) = e^{-W+L} u^{++\M}~.
  \end{equation}
Contracting both sides of these two equations in
indices $\M$, we get the expression for the field $W$
in terms of chiral harmonics
  \begin{equation}\label{Wsol}
  e^{-2W} = \frac{1}{2} l^{(--)\M} r^{(++)}_{\M}~.
  \end{equation}
For $D=3$  Eq. \p{Wsol} produces the general
solution of the Liouville equation in a special parametrization
(see Appendix ).

A similar solution for the Liouville equation in the conformal
gauge was presented in Ref. \cite{Kapustn}. The Cartan-Penrose
representation in terms of bosonic spinors was
used there for chiral light-like vectors $l^{(--)\M}$ and
$r^{(++)}_{\M}$.

In the generic case of higher $D$ it is necessary to get the suitable
representation for the $SO(D-2)$ matrices $G^{ij}$ as well.

\subsection{Relation of general and chiral harmonics:
solving the \break sigma--model--type equation}

To obtain the expression for $SO(D-2)$ matrix
field $G$, let us analyze the consequences of Eq. \p{rl=u} for the
moving frame vectors  $u^i$.
First of all, one finds
 \begin{equation}
\label{l--uir++ui}
 l^{(--)\underline{m}} u^i_{\underline{m}} = 0 , \qquad
 r^{(++)\underline{m}} u^i_{\underline{m}} = 0~. \qquad
\end{equation}
Eqs. \p{l--uir++ui} mean that the decompositions of  the
$u^i$ harmonic over the
chiral left- and right-moving ones  have no terms proportional
to $l^{(++)}$ and $r^{(--)}$, respectively
 \begin{equation}
\label{uidecl}
u^i_{\underline{m}}
= - (l^j_{\underline{m}} - V_{(--)}^j
l^{(--)}_{\underline{m}}) U^{ji}_{{\cal L}}~,
\qquad
\end{equation}
 \begin{equation}
\label{uidecr}
u^i_{\underline{m}} = - (r^j_{\underline{m}}
+  V_{(++)}^i r^{(++)}_{\underline{m}}) U^{ji}_{{\cal R}}~.
\end{equation}

The newly introduced matrices $U^{ji}_{{\cal L}}$ and $U^{ji}_{{\cal L}}$
are expressed through the contractions of chiral harmonics with the
generic ones as follows
  \begin{equation}
\label{ULUR}
U^{ji}_{{\cal L}}
=
l^{j\underline{m}}
u^i_{\underline{m}}~,
\qquad
U^{ji}_{{\cal R}}
=  r^{j\underline{m}}
u^i_{\underline{m}}~.
\end{equation}
So they can easily be checked to be orthogonal matrices
  \begin{equation}
\label{UU=I}
U^{jk}_{{\cal L}}
U^{ik}_{{\cal L}}
=  \delta^{ji}~,
\qquad
U^{jk}_{{\cal L}}
U^{ik}_{{\cal L}}
=  \delta^{ji}~.
\end{equation}
The proof uses the unity decomposition
\p{compl}
and the corollaries of Eqs. \p{rl=u}
  \begin{equation}
\label{uli}
 u^{--\underline{m}} l^i_{\underline{m}} = 0~, \qquad
 u^{++\underline{m}} r^i_{\underline{m}} = 0~.
\end{equation}

To find the relations between $U_{{\cal L},{\cal R}}$ and
$G_{{\cal R},{\cal L}}$, one should consider
the derivatives of the $U_{{\cal L},{\cal R}}$  fields
$$
\partial_{(--)}
U^{ji}_{{\cal L}}
=  l^{j\underline{m}}
\partial_{(--)} u^i_{\underline{m}}
\qquad \mbox{and} \qquad
\partial_{(++)}
 U^{ji}_{{\cal R}}
=  r^{j\underline{m}}
\partial_{(++)} u^i_{\underline{m}}
$$
and use the decomposition \p{hdif} to express the derivatives
of the generic harmonics in terms of components of the Cartan forms
\p{fdxi+}--\p{aijhol}.
In such a way  one arrives at

 \begin{equation}
\label{dUdG}
\left(\partial_{(--)}
U^{kj}_{{\cal L}}\right) U^{ki}_{{\cal L}}
= \left(\partial_{(--)}
G^{jk}_{{\cal R}}\right) G^{ik}_{{\cal R}}~, \qquad
\left(\partial_{(++)}
U^{kj}_{{\cal R}}\right) U^{ki}_{{\cal R}}
= \left(\partial_{(++)}
G^{jk}_{{\cal L}}\right) G^{ik}_{{\cal L}}~.
\end{equation}
Eqs. \p{dUdG} mean that the $G_{\cal L}$ ($G_{\cal R}$)
field differs from the
 (transposed)  matrix field
$U_{\cal R}$ ($U_{\cal L}$)
by an affine $SO(D-2)_L$ ($SO(D-2)_R$ )
transformation only (cf. \p{SO2})
   \begin{equation}
\label{G=HU}
G^{ij}_{{\cal R}} =
H_L^{jk}(\xi^{(++)})
U^{ki}_{{\cal L}}
=
u^{i\underline{m}}
l^k_{\underline{m}}
H_L^{kj}(\xi^{(++)})~, \qquad
\end{equation}
$$
G^{ij}_{{\cal L}} =
H_R^{jk}(\xi^{(--)})
U^{ki}_{{\cal R}}=
u^{i\underline{m}}
r^k_{\underline{m}}
H_R^{jk}(\xi^{(--)})~,
 \qquad
$$
  \begin{equation}
\label{HLHR}
H_L H_L^T = I , ~~~ \partial_{(--)} H_L = 0~, \qquad
H_R H_R^T = I , ~~~ \partial_{(++)} H_R = 0~.
\end{equation}

The expression for $V_{(++)}^i$ follows from
 the first equation in \p{l--uir++ui} upon substituting
 \p{uidecr} and using Eqs.\p{rl=u}, \p{G=HU}, \p{Wsol}.
In this way one gets
  \begin{equation}
\label{V++i}
V_{(++)}^i=  {1 \over 2} e^{2W}
l^{(--)\underline{m}}
r^i_{\underline{m}}~.
\end{equation}
In the same manner one can obtain
  \begin{equation}
\label{V--i}
V_{(--)}^i=   {1 \over 2} e^{2W}
r^{(++)\underline{m}}
l^i_{\underline{m}}
\end{equation}
from Eq. \p{uidecl} and the second of Eqs. \p{l--uir++ui}.

Now we can rewrite Eqs. \p{uidecl}, \p{uidecr} in terms of
chiral harmonics and the functions entering the nonlinear equations
\p{Liouville}, \p{Smodel}
 \begin{equation}
\label{uidecl1}
u^i_{\underline{m}}
=  G^{ik}_{{\cal R}} H^{kj}_{L}
(-  l^j_{\underline{m}}
+ e^{2W}
r^{(++)\underline{n}}
l^j_{\underline{n}}
l^{(--)}_{\underline{m}} )~,
\qquad
\end{equation}
 \begin{equation}
\label{uidecr1}
u^i_{\underline{m}}
=  G^{ik}_{{\cal L}} H^{kj}_{R}
(-  r^j_{\underline{m}}
+ e^{2W}
l^{(--)\underline{n}}
r^j_{\underline{n}}
r^{(++)}_{\underline{m}} )~.
\end{equation}

{}From Eqs. \p{uidecl1} and \p{uidecr1}
we obtain the expression
   \begin{equation}
\label{GLTGR}
(G^T_{{\cal L}}G_{{\cal R}})^{ij} =
 H^{ik}_{L}
(-  l^{k\underline{m}} r^l_{\underline{m}}
+ {2(r^{(++)\underline{n}} l^k_{\underline{n}})
(l^{(--)\underline{n'}} r^l_{\underline{n'}})
\over
(l^{(--)\underline{n''}} r^{(++)}_{\underline{n''}}) } )
  H^{jl}_{R}~,
\end{equation}
which provides the general solution for the sigma--model--like
equation \p{WZNW1} in the gauge \p{gauge}
$$
G_{{\cal R}}^{ij} = \delta^{ij}~, \qquad
G_{{\cal L}}^{ij} \equiv G^{ij}~,
$$
   \begin{equation}
\label{Gsol}
G^{ij} =
 H^{ik}_{L}
(-  l^{k\underline{m}} r^l_{\underline{m}}
+ {2(r^{(++)\underline{n}} l^k_{\underline{n}})
(l^{(--)\underline{n'}} r^l_{\underline{n'}})
\over
(l^{(--)\underline{n''}} r^{(++)}_{\underline{n''}}) } )
  H^{jl}_{R}~.
\end{equation}

To complete the description of the general solution of the string-inspired
system of nonlinear equations \p{Liouville0}, \p{WZNW1}, we have to present
the
expressions for chiral vector fields $M_{(\pm\pm)}^{(\mp\mp )i}$
\p{chirality}
in terms of the chiral harmonics.
This can be easily done by applying the derivatives
$\partial_{(++)}$
and
$\partial_{(--)}$
to both sides of Eqs. \p{uidecl1}, \p{uidecr1} and contracting the results
with the vectors $l^{(++)}$and $r^{(--)}$, respectively.
Then, using Eqs. \p{fdxi+}, \p{fdxi-},
one obtains
\begin{equation}
\label{M--exp}
M^{(++)i}_{(--)} = -  H_R^{ij}
r^{(++)\underline{m}}
\partial_{(--)}
r^j_{\underline{m}}~,
\end{equation}
\begin{equation}
\label{M++exp}
M^{(--)i}_{(++)}
= -  H_L^{ij}
l^{(--)\underline{m}} \partial_{(++)} l^j_{\underline{m}}~.
\end{equation}

Thus the chiral vectors $M^{(++)i}_{(--)}$ and $ M^{(--)i}_{(++)}$
appearing in Eqs. \p{Liouville0}, \p{WZNW1}
coincide with the covariant components
$f_{(\pm\pm)}^{(\mp\mp )i}$ \p{f++R}, \p{f--L}
of the chiral Cartan forms
which form a basis on the ``chiral spheres'' $S^{D-2}_R$
and $S^{D-2}_L$, respectively.

\subsubsection{General solution of the string--inspired nonlinear equations}

Eqs. \p{Wsol}, \p{Gsol}, \p{M--exp} and \p{M++exp}
provide
{\sl the general solution} for the
system of nonlinear equations \p{Liouville0}, \p{WZNW1}, \p{chirality}.

To be convinced of this, one has to take into account the following.
\begin{itemize}
\item
When obtaining the expressions \p{Wsol}, \p{Gsol}, \p{M--exp},
\p{M++exp} for all the functions which enter Eqs. \p{Liouville0}, \p{WZNW1},
\p{chirality}, we started from the general solution of the bosonic
string equations of motion in the standard Nambu-Goto approach
and then used these solutions in the equations of the geometric
approach.
\item
As we demonstrated in Section 1, the equations of the geometric approach
describing the extrinsic geometry of the D--dimensional  bosonic string
worldsheet uniquely produce the system of nonlinear equations
\p{Liouville0}, \p{WZSmodel}, \p{chirality}.
\item The equations of geometric approach and, therefore,
Eqs. \p{Liouville0}, \p{WZNW1}, \p{chirality}
specify the bosonic string worldsheet uniquely
(up to symmetry transformations, see e.g. \cite{Ei}), and thus
describe exactly the same dynamical system as the ordinary
(linear) string equations of motion.
\item
The general solution \p{Wsol}, \p{Gsol}, \p{M--exp}, \p{M++exp}
is written in terms of two sets of chiral spinor
harmonics \p{chiharm}, which
parametrize two copies of the compact coset \p{sphere}
 $$S^{D-2}
={SO(1,D-1) \over SO(1,1) \times SO(D-2)~ \semiprod ~K_{D-2}} $$
isomorphic to the sphere $S^{D-2}$. Thus it contains
$(D-2)$ right--moving and $(D-2)$ left--moving  degrees of freedom, the
same number as that of independent degrees of freedom
of the general solution \p{strsol} of the
standard  string equations of motion \p{str0}.

\end{itemize}

With this reasoning in mind, we conclude that the expressions
\p{Wsol}, \p{Gsol}, \p{M--exp} and \p{M++exp}
 for the functions $W$, $G$ and
$M^{(--)i}_{(++)}, M^{(++)i}_{(--)}$ obtained from the general solution
of the standard string equations {\sl have to provide the general solution}
of the geometric approach equations Eqs. \p{Liouville0}, \p{WZNW1},
\p{chirality} (up to superfluous symmetry transformations).

\section{On the group theoretical and geometrical structure of the solution}

In the course of deriving the general solution
\p{Wsol}, \p{Gsol}, \p{M--exp}, \p{M++exp}
we have got the expressions for the moving frame vectors
\p{harm} in terms of
chiral harmonics \footnote{More precisely, we have got
the first two equations in each  set \p{mapR}, \p{mapL}
while the third one
can be restored using the orthonormality conditions  \p{ortho}, \p{orthoR},
\p{orthoL}. }

 \begin{eqnarray}\label{mapR}
  u^{++}_{\M}&=& e^{W-L} r^{(++)}_{\M}~, \nonumber \\
  u^{i}_{\M}  &=&
    -(\GL H_R)^{ij} (r^{j}_{\M} -  V_{(++)}^{~~~j} r^{(++)}_{\M})~, \\
  u^{--}_{\M} &=& e^{-(W-L)} (r^{(--)}_{\M}
+ r^{(++)}_{\M} V_{(++)}^{~~~j} V_{(++)}^{~~~j} -
    2 r^{i}_{\M} V_{(++)}^{~~~i} )~,
 \nonumber
  \end{eqnarray}
  \begin{eqnarray}\label{mapL}
\nonumber
  u^{++}_{\M} &=& e^{-(W+L)} (l^{(++)}_{\M}
+ l^{(--)}_{\M} V_{(--)}^{~~~j} V_{(--)}^{~~~j} -
    2 l^{i}_{\M} V_{(--)}^{~~~i} )~,
\\
  u^{i}_{\M}  &=&
    -(\GR H_L)^{ij} (l^{j}_{\M} -  V_{(--)}^{~~~j} r^{(--)}_{\M})~, \\
 u^{--}_{\M}&=& e^{W+L} r^{(--)}_{\M}~ . 
 \nonumber
  \end{eqnarray}

If the functions $W$ and ${G}_{\cal L,R}$ satisfy the
 nonlinear equations \p{Liouville0}, \p{WZNW1},
then Eqs. \p{mapR}, \p{mapL} can be regarded as the  solution of the
corresponding associated linear system defined by Eqs.
\p{hdif}, with the Cartan forms \p{pC} being specified
by Eqs.  \p{fdxi+}, \p{fdxi-}, \p{omhol}, \p{aijhol}).

The solution of the zero curvature representation given by
the Maurer-Cartan equations \p{MC} with the Cartan forms
\p{pC} can be obtained by differentiating \p{mapR}, \p{mapL}.
The solution is presented by the following expressions for
the generic Cartan 1-forms \p{+i}--\p{ij} in terms of chiral once
\p{f++R}--\p{AijR} and \p{f++L}--\p{AijL}:

  \begin{equation}\label{f++fR}
  f^{++i} = e^{W-L} (\GL H_R)^{ij} f^{(++)j}_R~,
  \end{equation}
  \begin{equation}\label{f--fR}
  f^{--i} = -e^{-(W-L)} (\GL H_R)^{ij}
\Big(f^{(--)j}_R - 2 {\cal D}_R V_{(++)}^{~~~j}
- 2 f^{(++)k}_R (V_{(++)}^{~~~k} V_{(++)}^{~~~j} -
{1 \over 2} \d^{kj} V_{(++)}^{~~~} V_{(++)}^{~~~l})\Big)~,
\end{equation}
  $$
 {\cal D}_RV_{(++)}^{~~~j} \equiv d V_{(++)}^{~~~j} + \om_R V_{(++)}^{~~~j} -
V_{(++)}^{~~~k} A^{kj}_R~,
 $$

  \begin{equation}\label{omomR}
  \om = \om_R + f^{(++)i} V_{(++)}^{~~~i} + d(W-L)~,
  \end{equation}
  \begin{equation}\label{aijaR}
  A^{ij} =  (\GL H_R)^{ik} (\GL H_R)^{jl}   \Big(A^{kl}_R
+ (\GL H_R)^{-1} d(\GL H_R ))^{kl}
          + 2 f^{(++)[k} V^{--l]}\Big)~,
  \end{equation}

  \begin{equation}\label{f--fL}
  f^{--i} = e^{W+L} (\GR H_L)^{ij} f^{(--)j}_L~,
  \end{equation}
  \begin{equation}\label{f++fL}
  f^{++i} = -e^{-(W-L)} (\GR H_L)^{ij}
\Big(f^{(++)j}_L - 2 {\cal D}_R V_{(--)}^{~~~j}
- 2 f^{(--)k}_L (V_{(--)}^{~~~k} V_{(--)}^{~~~j} -
{1 \over 2} \d^{kj} V_{(--)}^{~~~} V_{(--)}^{~~~l})\Big)~,
\end{equation}
  $$
 {\cal D}_LV_{(--)}^{~~~j} \equiv d V_{(--)}^{~~~j} + \om_R V_{(--)}^{~~~j} -
V_{(--)}^{~~~k} A^{kj}_L~,
 $$

  \begin{equation}\label{omomL}
  \om = \om_R + f^{(--)i} V_{(--)}^{~~~i} - d(W+L)~,
  \end{equation}
  \begin{equation}\label{aijaL}
  A^{ij} =  (\GL H_R)^{ik} (\GL H_R)^{jl}   \Big(A^{kl}_R
+ (\GL H_R)^{-1} d(\GL H_R ))^{kl}
          + 2 f^{++[k} V^{--l]}\Big)~.
  \end{equation}

\bigskip

On the other hand, as follows from the consideration in
the previous section, the explicit form of the general solution
\p{Wsol}, \p{Gsol}, as well as the expressions
 \p{V++i},  \p{V--i} for the 'boost' parameters,
can be obtained algebraically from Eqs. \p{mapL}, \p{mapR}
(with making use of the orthonormality constraints \p{ortho},
\p{orthoR}, \p{orthoL} for the generic and chiral moving frame harmonics).

\bigskip

An intriguing point is that Eqs. \p{mapL}, \p{mapR} generating
the general solution {\sl have the form
of the parabolic symmetry transformations}
\p{tmapL}, \p{tmapR} of the chiral harmonics, but with non-chiral parameters.

\bigskip

Thus the prescription of how to solve the nonlinear equations
\p{WZNW1}, \p{Liouville0} can be formulated as follows.

Let us introduce the two set of chiral harmonics \p{chiharm}, \p{orthoR},
\p{orthoR} which
 define maps of the right  (left) light-cone sectors of the worldsheet
${\cal M}^{(0,1)} \equiv \{ (\xi^{(--)}) \}$
(${\cal M}^{(1,0)} \equiv \{ (\xi^{(++)}) \}$) onto the sphere $S^{D-2}$
$$
r^{(\A )}_{\M }~:~~~
{\cal M}^{(0,1)} \equiv \{ (\xi^{(--)}) \}
~~ \rightarrow ~~S^{D-2}={SO(1,D-1) \over SO(1,1) \times SO(D-2)
\semiprod K_{D-2}}~,$$
$$
r^{(\A )}_{\M }~:~~~
{\cal M}^{(1,0)} \equiv \{ (\xi^{(++)}) \}
~~ \rightarrow ~~
S^{D-2}={SO(1,D-1) \over SO(1,1) \times SO(D-2)
\semiprod K_{D-2}}~.$$
Further, let us assume that the generic harmonics \p{harm}, \p{ortho}
$$
u^{\A }_{\M }~:~~~
{\cal M}^{(1,1)}
~~ \rightarrow ~~
{SO(1,D-1) \over SO(1,1) \times SO(D-2)}$$
are related to the chiral ones by the parabolic transformations
\p{mapR}, \p{mapL} (with the chiral parameters $H_L$ and $H_R$
omitted for simplicity).

Then let us exploit the $SO(D-2)$ gauge freedom
to fix the $SO(D-2)$ rotation matrix in \p{mapR}
to be the unity one. The $SO(D-2)$ rotation matrix in \p{mapL}
taken in this gauge
provides us with the solution \p{Gsol}
of the WZNW sigma--model-type equation
\p{WZNW1} (with the chiral vectors ${M}^{(++)i}_{(--)}$,
${M}^{(++)i}_{(--)}$ determined by
the homogeneously transforming components of chiral Cartan
forms \p{f++R}, \p{f--L}
$$
M^{(++)i}_{(--)}= f^{~(++)i}_{(--)R}~, \qquad
M^{(++)i}_{(--)}=f^{~(--)i}_{(++)L}
$$
(cf. \p{M++exp}, \p{M--exp})).
The product of the $SO(1,1)$ transformation factors
from \p{mapR} and \p{mapR} produces the general solution of the
Liouville--like equation \p{Liouville0}.

\bigskip

Since the parabolic transformations \p{mapL}, \p{mapR} have a
``triangular'' form, one can expect that the described method of
getting the general solution bears a tight relation to the known
group--theoretical methods of solving nonlinear equations, like
those developed and used in \cite{Leznov,IK1,Ivanov}.
A detailed examination of such a relationship
could provide a deeper insight into the nature of integrability
and we consider it as an interesting problem for the future study.

\section{Conclusion}

Thus we have obtained the general solution of the string--inspired
nonlinear equations \p{WZNW1}, \p{Liouville}, \p{chirality}
describing the extrinsic geometry of the bosonic string worldsheet
in the geometrical approach \cite{geom,BNbook,bpstv,zero}.

The solution is given in terms of the two sets of chiral (left--moving
and right--moving) Lorentz harmonic variables \p{chiharm}, \p{orthoR},
\p{orthoL} and has the form (after fixing a gauge with respect to some
extra symmetries)
  \begin{equation}\label{W=rl}
  e^{-2W} = \frac{1}{2} r^{(++)}_{\M} l^{(--)M}~,
  \end{equation}
  \begin{equation}\label{G=lr}
  G^{kj}  = -l^{k}_{\M} r^{j\M} +
    \frac{r^{(++)}_{\M} l^{k\M} l^{(--)}_{\N} r^{j\N}}
         {r^{(++)}_{\underline p} l^{(--)\underline p}}~,
  \end{equation}
  \begin{equation}\label{M+=lr}
  M_{(--)}^{(++)i}(\km)
= r^{(++)\M} \pam r^{i}_{\M}~,
  \end{equation}
  \begin{equation}\label{M-=lr}
    M_{(++)}^{(--)i}(\kp) = l^{(--)\M} \pap l^{i}_{\M}~.
  \end{equation}

The analysis of the solution of the associated linear system demonstrates
that the general solution we have found can be regarded as the
parabolic subgroup
${SO(1,1) \times SO(D-2)
\semiprod K_{D-2}}$
transformations \cite{gds} of the chiral harmonics with non-chiral parameters.
As these transformations are of a ``triangle'' form in the matrix
representation, we can expect a close relation of our approach to
the existing group--theoretical methods of solving nonlinear equations
\cite{Leznov,IK1,Ivanov}.
It is an interesting task for further study to elaborate such a relation
in more detail.

A natural direction of extending our results is to look for
the solution of a supersymmetric generalization of the
considered nonlinear equations. Such a system describes the extrinsic
geometry of the worldsheet superspace of $D=3,4,6,10$ superstring models
\footnote{Let us note that the  explicit form of
such supersymmetric equations are known at present for
the $D=3,~N=1,2$  cases   \cite{bsv1} only.}.

\bigskip

\noindent{\bf Acknowledgments}

\vspace{0.3cm}
\noindent The authors are grateful to D. Sorokin, S. Krivonos, B. Julia, 
F. Toppan 
for interest to this work   and useful remarks. One of the author (I.B.) 
would like to thank
Prof. M. Virasoro and Prof. S. Randjbar--Daemi for the hospitality at the
Abdus Salam ICTP, where the part of this work was done.
The work was supported in part 
by the INTAS Grants {\bf 96-0308,
96-0538, 93-127ext}, the RFBR Grant {\bf 96-02-17634} 
and by the Fonds zur F\"{o}rderung der 
wissenschaftlichen Forschung under Project {\bf No M472--TPH} (IB).

\bigskip

\section*{Appendix : Relation with the standard form of general solution of
the Liouville equation}

Here we demonstrate that for $D=3$ case Eq. \p{Wsol}
reproduces the general solution of the nonlinear Liouville
equation.
Using the well-known
parametrization of the $SO(1,2)$ matrices
  \begin{eqnarray}\label{a1}
  l^{\M}_{(\A )} &=& \left(\begin{array}{ccc}
    \cosh{A_L} & \sinh{A_L} & 0 \\
    \sinh{A_L} & \cosh{A_L} & 0 \\
    0 & 0 & 1
    \end{array}\right)~, \nonumber \\
  r^{\M}_{(\A )} &=& \left(\begin{array}{ccc}
     \cosh{A_R} & -\sinh{A_R} & 0 \\
    -\sinh{A_R} & \cosh{A_R}  & 0 \\
    0 & 0 & 1 \end{array}\right)~,
  \end{eqnarray}
  $$
  \pam A_L = \pap A_R = 0~, \nonumber
  $$
one finds the following form of chiral harmonics
\p{chiharm}
  \begin{eqnarray}\label{a2}
  l^{(--)\M} &=&  \left(\begin{array}{ccc}
    \cosh{A_L}~, & \sinh{A_L}~, & 1 \end{array}\right)~, \nonumber \\
  l^{\perp \M}  &=&  \left(\begin{array}{ccc}
    \sinh{A_L}~, & \cosh{A_L}~, & 0 \end{array}\right)~, \\
  l^{(++)\M} &=&  \left(\begin{array}{ccc}
    \cosh{A_L}~, & \sinh{A_L}~, &-1 \end{array}\right)~, \nonumber \\
  r^{(--)\M} &=&  \left(\begin{array}{ccc}
    \cosh{A_R}~, & -\sinh{A_R}~,& 1 \end{array}\right)~, \nonumber \\
  r^{\perp \M}  &=&  \left(\begin{array}{ccc}
    -\sinh{A_R}~,& \cosh{A_R}~, & 0 \end{array}\right)~, \\
  r^{(++)\M} &=&  \left(\begin{array}{ccc}
    \cosh{A_R}~, & -\sinh{A_R},&-1 \end{array}\right)~. \nonumber
  \end{eqnarray}
Substituting these expressions into the Eqs. \p{Wsol}, \p{M--exp},
\p{M++exp} with $D=3$, one gets
  \begin{equation}\label{a3}
  e^{-W} = \cosh \frac{A_L + A_R}{2}~,
  \end{equation}
  \begin{equation}\label{a4}
  \Mpm{\perp }(\km) = r^{(++)\M} \pam r^{\perp }_{\M} = - \pam A_R~,
  \end{equation}
  \begin{equation}\label{a5}
  \Mmp{\perp }(\kp) = l^{(--)\M} \pap l^{\perp }_{\M} = \pap A_L~.
  \end{equation}

The relation to the standard parametrization of the general solution of
the Liouville equation (see e.g. \cite{kulish}) is given by
$$
f_{L,R}=\exp A_{L,R}~.
$$

\end{document}